\documentclass[fleqn,10pt]{wlscirep}

\usepackage[utf8]{inputenc}
\usepackage[T1]{fontenc}
\usepackage{upgreek}
\usepackage{siunitx}
\usepackage{libertine} 
\usepackage{graphicx}
\usepackage{dcolumn}
\usepackage{bm}
\usepackage{etoolbox}
\usepackage{color}
\usepackage{subfig}
\usepackage{floatrow}

\usepackage{textgreek} 
\newcommand{\um}{\textmu m} 
\newcommand{\topowmath}[2][0]{\ifnumcomp{0}{=}{#1}{}{#1~\cdot~}10^{#2}}
\newcommand{\topow}[2][0]{\ifnumcomp{0}{=}{#1}{}{$#1~\cdot$~}$10^{#2}$}

\newcommand{\Wcm}[2][0]{\topow[#1]{#2}~W/cm$^2$} 
\newcommand{\invcc}[2][0]{\topow[#1]{#2}~cc$^{-1}$}
\newcommand{\invccmath}[2][0]{\topowmath[#1]{#2}~\text{cc}^{-1}}

\begin{document}



\title{Experimental observation and computational modeling of radial Weibel instability in high intensity laser-plasma interactions}

\author[1,5,*]{Gregory K. Ngirmang}
\author[2]{John T. Morrison}
\author[2]{Kevin M. George}
\author[3]{Joseph R. Smith}
\author[2]{Kyle D. Frische}
\author[3]{Chris Orban}
\author[3,4]{Enam A. Chowdhury}
\author[5]{W. Mel Roquemore}

\affil[1]{National Academies of Sciences, Engineering, and Medicine, Washington DC, USA}
\affil[2]{Innovative Scientific Solutions, Inc., Dayton, Ohio 45459, USA}
\affil[3]{Ohio State University, Department of Physics Columbus, Ohio 43210, USA}
\affil[4]{Intense Energy Solutions, LLC., Plain City, Ohio 43064, USA}
\affil[5]{Air Force Research Laboratory, Dayton, Ohio 45433, USA}
\affil[*]{ngirmang.1@osu.edu}



\begin{abstract}
When a relativistic intensity laser interacts with the surface of a solid density target, suprathermal electron currents are subject to Weibel instability filamentation when propagating through the thermal population of the bulk target. We present time resolved shadowgraphy of radial ionization front expansion and Weibel instability filamentation within a thin, sub-micron, sheet initiated by irradiation with a short pulse, high intensity laser. High temporal (100~fs) and spatial (\SI{1}{\micro\meter}) resolution shadowgraphy of the interaction reveals a relativistic expansion of the ionization front within a \SI{120}{\micro\meter} diameter region surrounding the laser-target interaction, corroborated by simulations to expand at $0.77c$, where $c$ is the speed of light. Filamentation within the patch persists for several picoseconds and seeds the eventual recombination and heating dynamics on the nanosecond timescale. Particle-in-cell simulations were conducted to elucidate the electron dynamics leading to the radial expansion of the critical surface. Computational results report the ionization expansion is due to field ionization of the expanding hot electron population. Filamentation within the expansion is due to the Weibel instability which is supported by the magnetic fields present.
  %

\end{abstract}

\flushbottom
\maketitle

\thispagestyle{empty}

%
\section*{Introduction} \label{introduction}

The development of chirped pulse amplification \cite{DonnaStricklandandGerardMorou1985} has lead to a dramatic increase in the intensity of ultrashort pulse laser systems. As a result, the study of extreme phenomena stemming from the interaction of relativistically intense lasers with matter has produced a rich field of research with broad and promising applications. Relativistic laser-plasma interactions (RLPI) generate exotic states of warm dense matter, accelerate charged particles to relativistic energies \cite{wilks2001energeticproton,TajimaDawson,morrison2015backward,Orban2015}, and drive fundamental plasma processes rarely observed in nature.

Examination of the RLPI event remains a challenge given the femtosecond-temporal and submicron-spatial scale evolution of the interaction. A variety of experimental techniques have been developed in an effort to better understand and diagnose these conditions~\cite{Haffa2019,Borghesi2003protonimage,Li2006protonimage,Liao2016protonradiographyB,Feister2014}. Nonetheless, the difficulty of resolving these temporal and spatial scales necessitate the use of computational particle-in-cell (PIC) simulations to model the RPLI physics of these scales directly. These codes which model RLPI with unprecedented detail are the only method to fully resolve these scales and play an important role explaining the underlying physics in these extreme conditions, although they represent an indirect probe of the experimental physics through corroboration with observation.

Utilizing a novel pump-probe imaging system, detailed in Feister~\textit{et~al.} 2014~\cite{Feister2014}, we generate time resolved movies of the RLPI with high temporal (100 femtosecond) and spatial (1 micron) resolution. With this technique, we observe the evolution of a hot electron induced, above critical density plasma expansion, radially oriented from the laser-target interaction region within a 450 nm thick liquid sheet target. Weibel instabilities form between the expanding hot electron population and neutralizing, cold, return current resulting in filamentation of this hot, dense plasma~\cite{Weibel1959}. Fully three-dimensional PIC modeling corroborate these findings with the presence and growth of strong, radially oriented, alternating magnetic fields within the target. 

Previous works have reported the occurrence of Weibel instabilities driven by RLPI generated hot electron populations, though solely in the longitudinal direction and predominantly with relevance to fast ignition configurations \cite{Sentoku2002,Sentoku2000,Wallace1987,Honda2000,Tatarakis2002,Califano2006,Tatarakis2003}. The implications of the radially oriented Weibel instability upon which we report is expected to be widespread given the common experimental conditions of a high-intensity laser incident onto a submicron thick planar target, but has yet to be accounted for in such studies. Non-uniform heating in the bulk target and imprinting of the filamentation pattern on lower energy target normal sheath accelerated ions are two actives areas of interest which is implicated in the impact of this work. Lastly, it is important highlight that this result was only made possible by the combination of several systems unique to the field of RLPI; short pulse pump-probe imaging scheme, high-repetition rate relativistically intense laser, transparent liquid target, and real time diagnostic feedback. 

\section*{Results}
\subsection*{Observation of Radial Filamentation}\label{sec:results}

\begin{figure}
  \begin{minipage}{.40\textwidth}
    \sidesubfloat[]{
      \includegraphics[width=\linewidth]{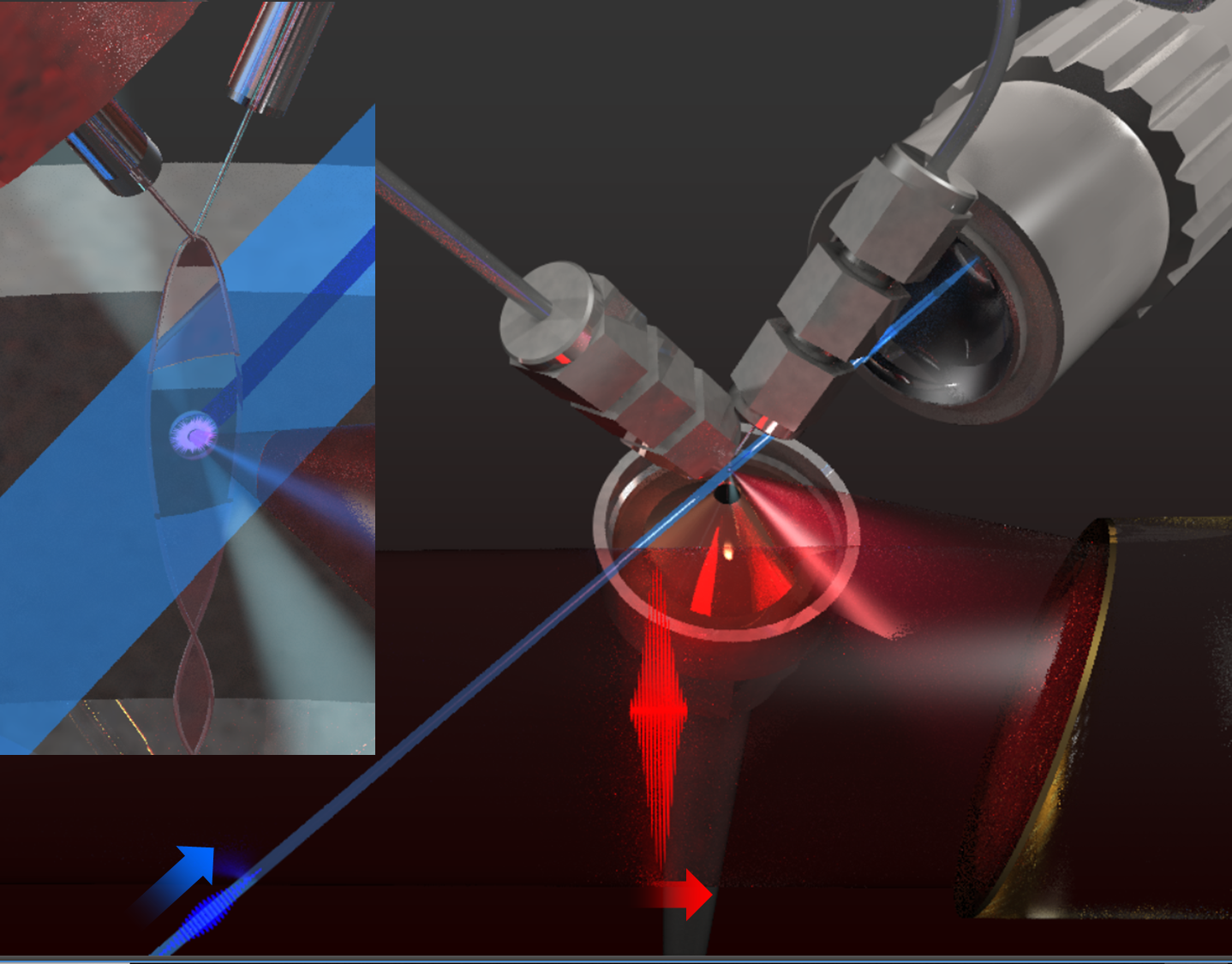}\label{fig:expsetup}
    }\\
    \vspace*{0.5cm}\\
    \sidesubfloat[]{
      \includegraphics[width=\linewidth]{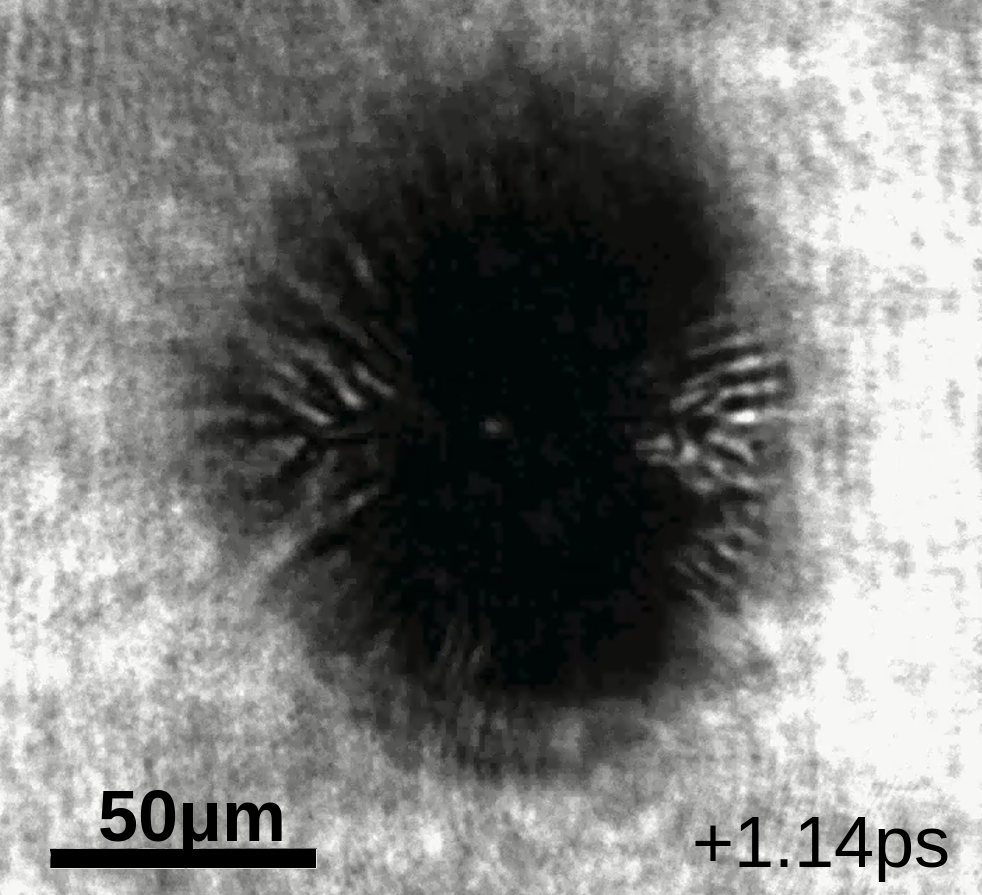}\label{fig:expfilaments}
    }
  \end{minipage}
  \hspace{0.05\textwidth}
  \begin{minipage}{.45\textwidth}
    \sidesubfloat[]{
      \includegraphics[width=\linewidth]{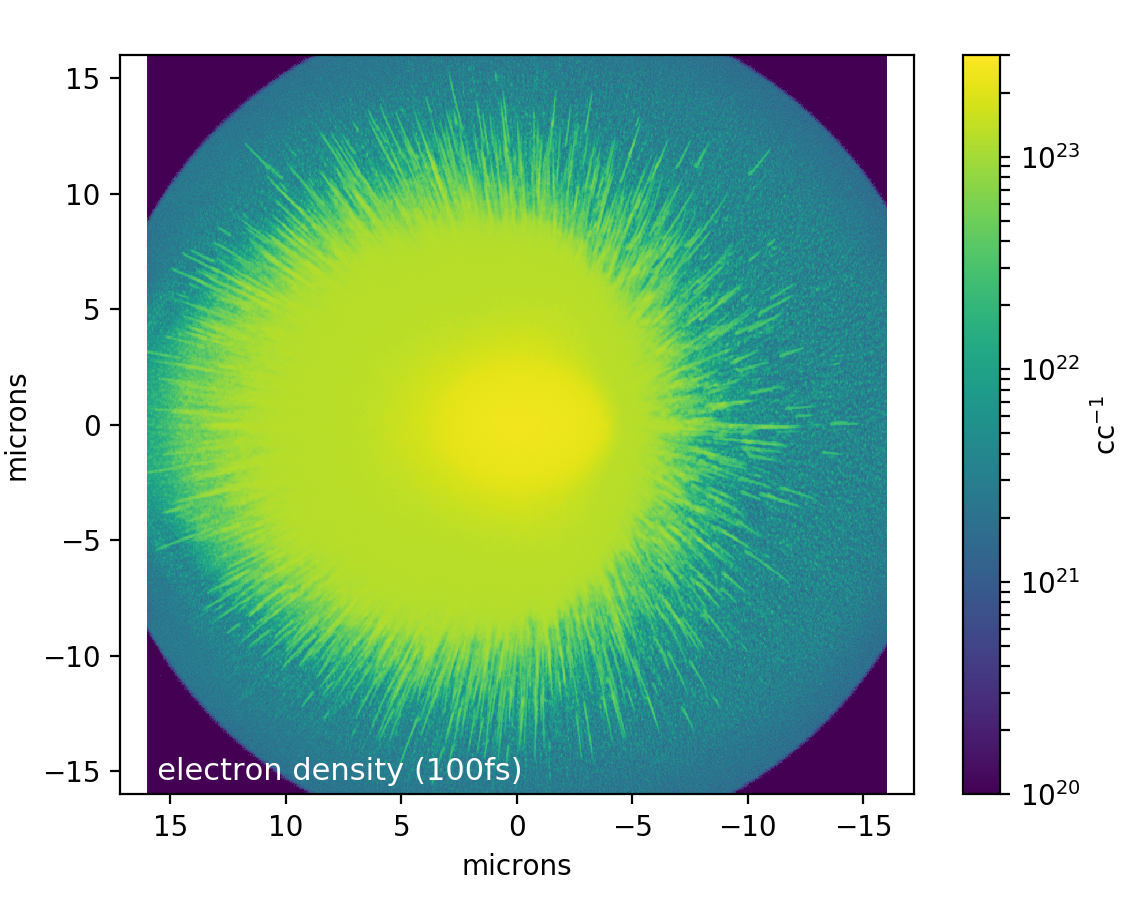}\label{fig:simelecdens}
    }\\
    \vspace*{0.1cm}\\
    \sidesubfloat[]{
      \includegraphics[width=\linewidth]{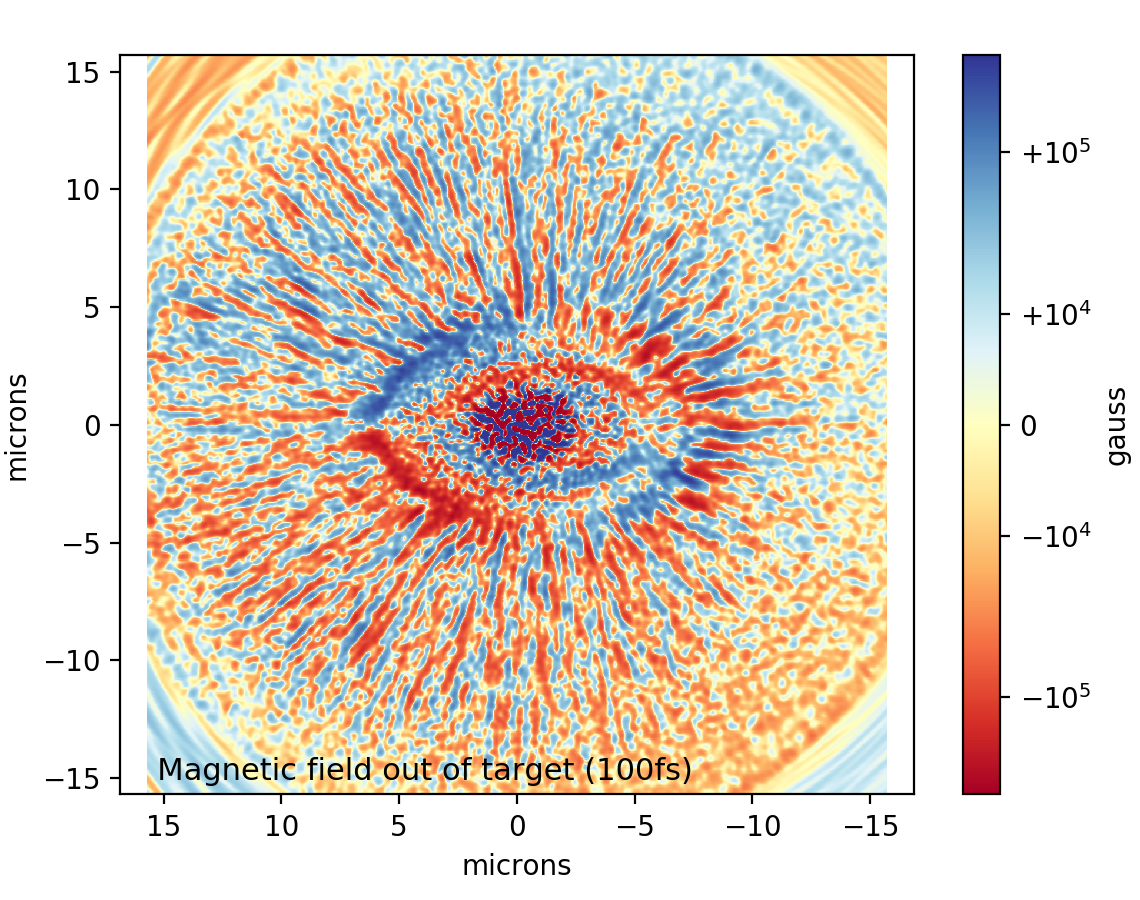}\label{fig:simBfield}
    }
    

  \end{minipage}%

  \caption{\protect\subref{fig:expsetup} Schematic of experimental setup employed to view the Weibel instability formation and evolution within a sub-micron thin, free flowing liquid sheet. A frequency doubled and shifted probe beam, synchronized to the main pulse with 420 nm wavelength, 80 fs pulse duration and variable delay was used to illuminate the sheet viewed with a 10x microscope objective. This system is capable of temporal resolution of 100 fs and arbitrary pump-probe delay. \protect\subref{fig:expfilaments} Frame of the shadowgraphy movie that demonstrates the filamentation feature. \protect\subref{fig:simelecdens} Particle-in-Cell free electron density illustrates the initial formation of the filamentation feature. \protect\subref{fig:simBfield} Magnetic field component out of the target plane in the Particle-in-Cell simulations demonstrating pinching around the filaments.}
  \label{fig:main}
\end{figure}

We present the observation of laser plasma surface expansion and the filamentation of this laser produced plasma in a sub-micron thick target, corroborated by the development of a radial Weibel instability in three-dimensional high resolution Particle-in-Cell (PIC) simulations. The experiment, displayed in Fig.~\ref{fig:expsetup}, was performed with a relativistic intensity, short pulse laser incident on a thin ($\sim$0.5\um{}) target~\cite{george2019target}. The laser-target interaction was illuminated by a short pulse probe beam created from frequency doubling different pulses of a common oscillator in a scheme described in the Feister et. al\cite{Feister2014} and in the Methods section. The use of a common oscillator allows the synchronization of the shadowgraphy with the main pulse. Utilizing the repetition rate, we develop time resolved movies of the target evolution, a frame of which is shown in Fig.~\ref{fig:expfilaments} and is supplied in the supplemental data. We interpret the darkening of the target as attenuation of the probe beam due to free electron density, and therefore serves as a measurement of the electron dynamics. On the outer edge of the dark spot, finger like features are present, suggesting that the electron density is filamented as it expands. We posit that as the ionization front moves radially away from the target, an electromagnetic instability results as the cold electron return current counter propagates. The instability we suggest is similar to the  Weibel instability given that the wavenumber of the filaments are normal to the direction of hot electron motion which is radially away from the laser spot. This perspective was substantiated by PIC simulations that modelled the initial laser target interaction. The free electron density in the target plane is found to filament femtoseconds after irradiation as shown in Fig.~\ref{fig:simelecdens}, and plots of the magnetic field component into and out of the target in Fig.~\ref{fig:simBfield} demonstrates a quasi-static pinching of the filaments, consistent with simulations of the Weibel instability in previous work in linear geometry (see for example, Fig.~5 of Tatarakis~\textit{et~al.}~\cite{Tatarakis2003}, Fig.~1 of Sentoku~\textit{et~al.}~\cite{Sentoku2002}, Fig.~4 of Wei~\textit{et~al.}~\cite{Wei2004}, Fig.~3 of G\"ode~\textit{et~al.}~\cite{Gode2017}). 

\floatsetup[figure]{style=plain,subcapbesideposition=top}

\subsection*{Temporal Shadowgraphy Movie}
\begin{figure*}[t]
	\includegraphics[width=\textwidth]{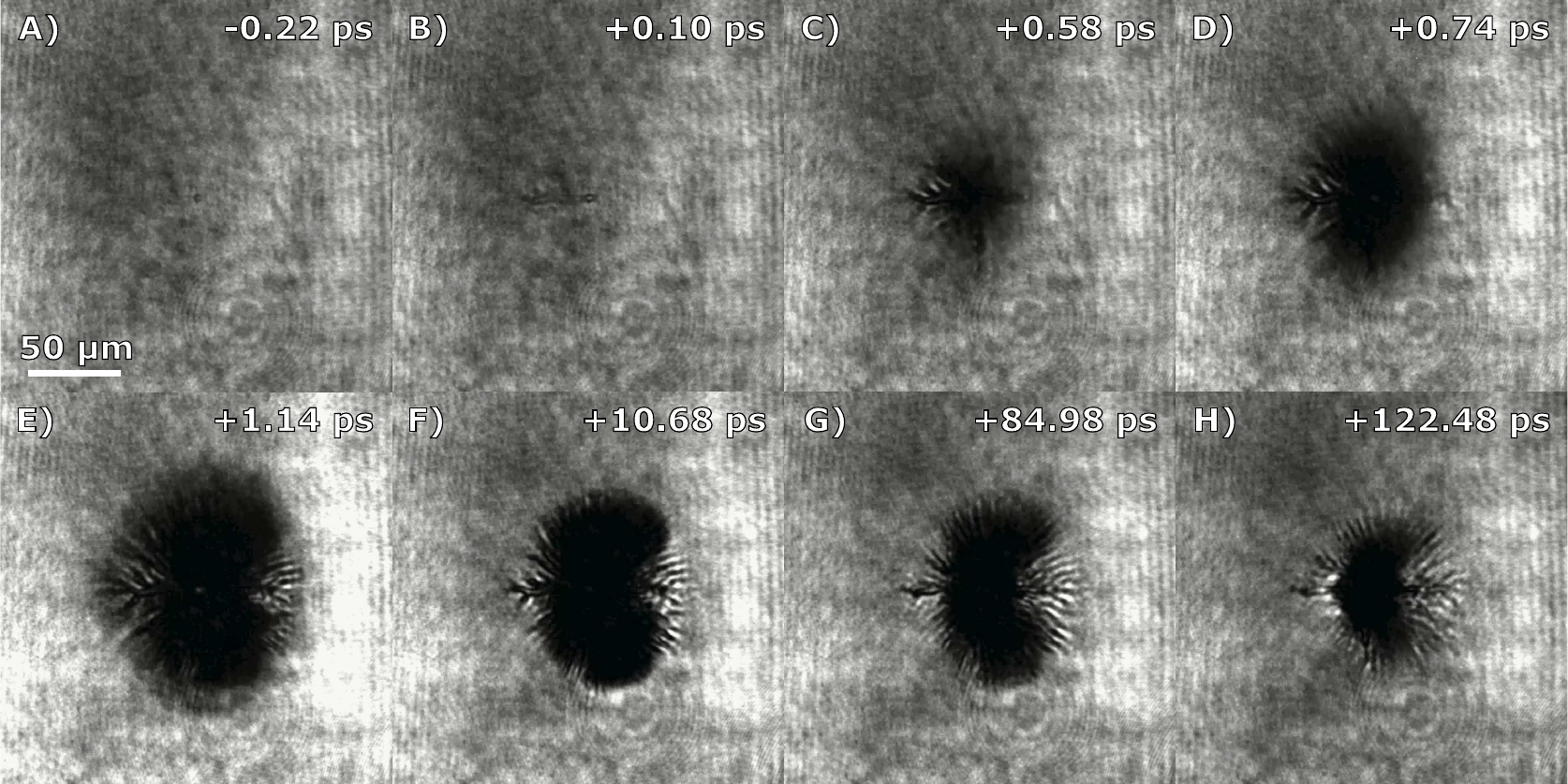}
	\caption{Short pulse probe beam shadowgraphic microscope images of the Weibel instability evolution from expanding energetic electrons within the thin liquid sheet. A) Preceeding the arrival of the main pulse, the target surface is perturbed by prepulse on the nanosecond timescale. B) A filament forms along the target surface in the direction of the laser pulse due to energetic electrons accelerated from coupling between the pulse and surface plasma waves. C) The rapidly expanding hot electron population induces field ionization in the cold electron population of the liquid leading to a growing dark ring which surrounds the laser point where the laser is incident onto the target. D) Filaments within the expanding ionization front grow due to anisotropy in the temperature distribution and are reinforced by magnetic fields initiated by the hot electron current. E) Growth of the ionization front stagnates after 1.14 ps. F) Recombination occurs as the plasma cools, shrinking the darkened patch. G) The plasma cools over a relatively long timescale (greater than 100 ps). H) Filaments persist at late times due to the high temperatures and electron densities caused by the Weibel instability.}
	\label{fig:movie}
\end{figure*}

Fig.~\ref{fig:movie} displays frames from the movies, showing the full time resolved nature of this measurement, particularly highlighting the evolution of the ionization front. The features develop as follows. First, in just hundreds of femtoseconds after the first features of darkening appears, a rapidly expanding filament like feature develops across the target in the direction of the $k$-vector along the target surface. Coincident with the filamentation is the initial expansion of a darkened region. After half a picosecond, the darkened patch grows and becomes apparent, with filaments also becoming discernible near the edges of the dark spot. We recommend that the reader view the video included in the supplement in order better understand the feature we observe and describe here.

It should be highlighted how the experimental observation of the dark feature and its evolution was facilitated by the time resolved nature of the movies. The diagnostic provides a means to observe the target evolution over a large range of timescales, from the picosecond to the nanosecond scale after the target interaction. This facilitated the discovery of the feature which would have required either foreknowledge or chance to discover. Moreover, the movie creation scheme detailed in the Methods section stitches together images from different main pulse shots into a single movie, and mere observation of the stability of a feature in the video allows an immediate gauge of the repetability and shot-to-shot stability of the feature in the experiment. The filamentary structures we observe are stable when viewed as a video, implying that the seed of these filaments is non-random and repeatable. This fact will be discussed in the following section. Finally, although not shown in Fig.~\ref{fig:movie} one can observe the later time hydrodynamic explosion of the target nanoseconds after the RLPI in the movies provided in the supplement, demonstrating the large temporal range provided by this technique.


\section*{Discussion}
\subsection*{Seeding of the Instability} \label{sec:seed}
\begin{figure}
\sidesubfloat[]{
    \includegraphics[width=0.35\textwidth]{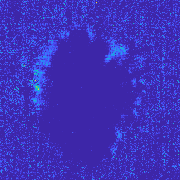}\label{fig:wirea}
} 
\sidesubfloat[]{
    \includegraphics[width=0.35\textwidth]{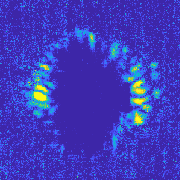}\label{fig:wireb}
}
\caption{Arbitrarily unit pseudo-color of the difference in sequential shadowgraphic frames at a fixed time delay during which the laser is directly perturbed by the use of a wire obstruction in the path of the laser before focusing, with yellow-to-green representing more variation. \protect\subref{fig:wirea} demonstrates the control case with no wire obstruction and \protect\subref{fig:wireb} shows the shadowgraphy variation when a wire is perturbing the mode.}
\label{fig:wire}
\end{figure}

As discussed previously, the filaments in these videos appear consistent in shape and darkness, and thus appear ``stable'' when viewed as a movie as can be observed in the supplementary video. Given the pump-probe method that takes snapshots across different shots, this means that the filaments are in fact the \emph{same} in configuration from shot-to-shot. This suggests that the instability process that leads to filamentation is a repeatable process seeded by an anisotropy that does not significantly vary across shots. Since the Weibel instability arises from an anisotropy in the momentum distribution of electrons, the shape of the laser focal spot at the target is the likely source. In order to test this, a movie was generated at a \emph{fixed} time delay of 1~ps when the filamentary structures are at time of maximum expansion, but over a set of shots in which the laser beam itself was perturbed by placing a 1 mm diameter wire into the laser beam before entering the target chamber. When the wire was placed in the beam, the radial filaments' angles were found to change as observed in the movie. It was also observed that motion of the wire caused the filaments to move, while if the wire was held obstructing but stationary, the filaments maintained their new orientations. Finally, the original filament structure was reestablished when the wire was removed. This leads us to conclude that the filaments are seeded by higher order terms in the laser mode which modify the laser focus at the target. Fig.~\ref{fig:wire} displays the variation in the filament structure observed in the movie and the actual movie generated is contained in the supplement.

\begin{figure}
\sidesubfloat[]{
    \includegraphics[width=0.30\textwidth]{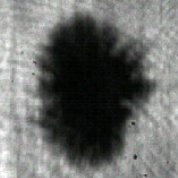}\label{fig:pola}
} 
\sidesubfloat[]{
    \includegraphics[width=0.30\textwidth]{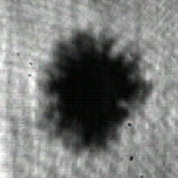}\label{fig:polb}
}
\sidesubfloat[]{
    \includegraphics[width=0.30\textwidth]{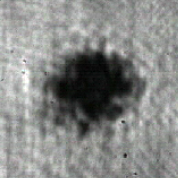}\label{fig:polc}
}
\caption{Shadowgraphic images with a 1 ps probe delay of experiments with identical laser parameters (see text) including energy and mode with incident \protect\subref{fig:pola} \emph{P}-polarization, \protect\subref{fig:polb} $45^{\circ}$, and \protect\subref{fig:polc} \emph{S}-polarization. The disparity in the extent of ionization between \emph{P} and \emph{S} polarization implies that the shape of the ionization feature is not exclusively dependent on the laser mode, and strongly depend on the laser target coupling. It should be noted that experiments on a different system, attributes the dependence to the laser mode shape\cite{Haffa2019}. }
\label{fig:polarization}
\end{figure}

 Polarization dependence of the filament structure was also studied. Movies were recorded with S, P, and circularly polarized light through the use of half and quarter waveplates placed in the beam before entry into the vacuum chamber. The ionization patches for linear polarization cases are shown in Fig.~\ref{fig:polarization}. The radial extent of the ionized region changes though the intensity is constant, so the ionized region cannot be exclusively attributed to radial intensity distribution of the laser focus. As the polarization was changed from P (maximum laser coupling) to S (minimum laser coupling) the size of the ionization patch shrunk and the filament length decreased though preserving a similar overall structure. These results indicate, for this experiment, the radial range of relativistic electron expansion is greater than the diameter at which the intensity of the focus is below the ionization threshold. We should note that previous work by Haffa \textit{et al.}~\cite{Haffa2019}. finds a dependence of the spatial extent of the ionization front is dependent on the mode specifically, in apparent contradiction to this result. One potential possibility is the relevant laser modes for Haffa \textit{et al.} generate relativistic electrons in the peak intensity which do not propagate beyond the region directly ionized by the laser.

\subsection*{Recombination}

About one picosecond after the first features on shadowgraphy appear, the ionization patch stops growing, and maintains its size for about 80 to 90 picoseconds, after which the darkened background begins to disappear. The disappearance occurs from the outer edges (further radial distance from the center) and inward, giving the appearance of shrinking. During this shrinking, the darkening decreases, making the filaments more apparent, although the edges of the filaments do also begin to vanish as well. This we interpret as occurring due to electron-ion recombination, and thus these movies serve as a probe of the electron-ion recombination rate for high density plasma. Finally 200 picoseconds after the initial laser interaction, the dark patch has reduced to a spot smaller than 30 microns, where it remains that size until the 300-400 picoseconds where the material begins to explode.

\subsection*{Simulations} \label{simulations}

The three-dimensional Particle-in-Cell (PIC) simulation using the Large-Scale Plasma (LSP) code~\cite{Welch2004} was employed in order to model this experiment and help elucidate the mechanism responsible for the filaments. Given the geometry of the experiment, lower dimensional simulations would not capture the full dynamics of the ionization correctly. This computationally expensive simulation was designed to closely follow the actual experiment, the details can be found in the Methods section. 

\begin{figure} 

\begin{minipage}{0.3\textwidth}
    \sidesubfloat[]{
        \includegraphics[width=\linewidth]{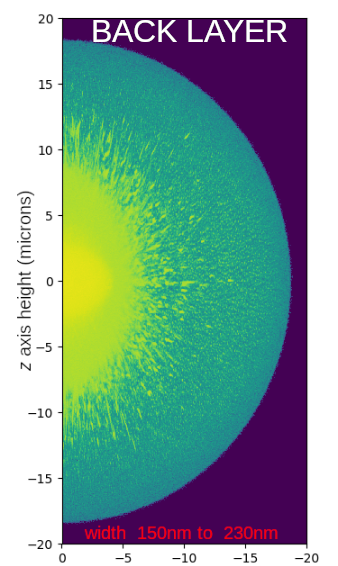}
        \label{fig:simbacklayer}
    }\\
    \sidesubfloat[]{
        \includegraphics[width=\linewidth]{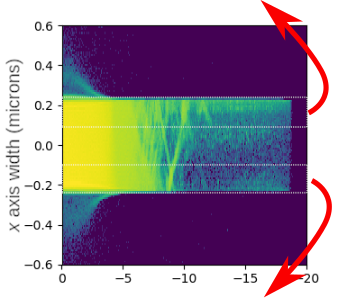}
        \label{fig:simdownxlayer}
    }\\
    \sidesubfloat[]{
        \includegraphics[width=\linewidth]{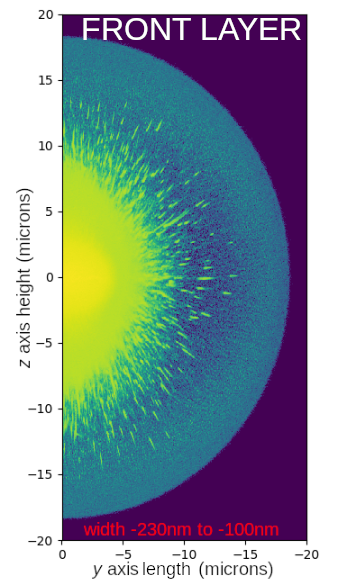}
    \label{fig:simfrontlayer}
    }
\end{minipage}
\hspace{0.05\textwidth}
\begin{minipage}{0.4\textwidth}

\sidesubfloat[]{  
  \includegraphics[width=\linewidth]{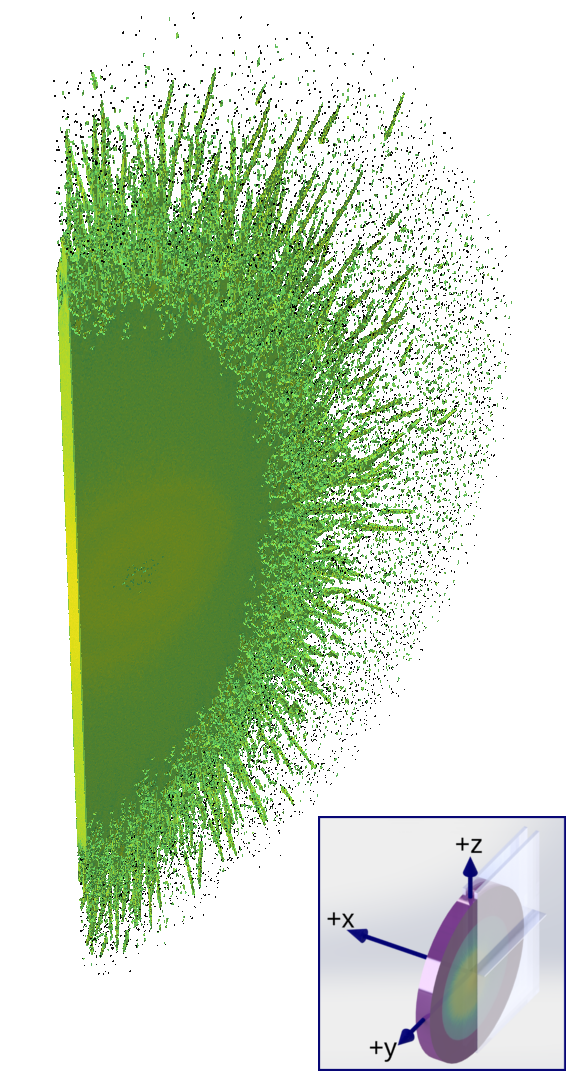}
  \label{fig:sim3d}
}
\\
\parbox[t][0.4\textheight]{1.1\linewidth}{
  \RawCaption{\caption{ Sub-figures \protect\subref{fig:simbacklayer} to \protect\subref{fig:simfrontlayer} show the free electron density variation through the thickness of the target, or the $x$-axis as shown in the inset. All figures consider the target density away from the laser forward direction, which is along the negative-$y$ axis. \protect\subref{fig:simdownxlayer} shows the free electron density averaged over half a micron in the transverse direction ($z$-axis), with the vertical dimension of the figure corresponding to the $x$-axis expanded to show detail. \protect\subref{fig:simbacklayer} shows the electron density on the back surface of the target, averaged 80~nm along the $x$-axis and \protect\subref{fig:simfrontlayer} shows the front of the target averaged through 130~nm along the $x$-axis. \protect\subref{fig:simdownxlayer} also shows, although distorted due to the expansion of vertical dimension, the random pitch of the filaments through the thickness axis of the target. \protect\subref{fig:sim3d} shows a volumetric three dimensional render of this half of the target which further highlights the orientation of the filaments. See the inset for convienience in understanding the orientation of these plots, in which the volumes displayed in \protect\subref{fig:simbacklayer} to \protect\subref{fig:simfrontlayer} are highlighted by the semi-transparent rectangles shown. The color map for these figures is the same as that for Fig.~\ref{fig:simelecdens}.}\label{fig:moresim}}}
\end{minipage}

\end{figure}

As discussed previously and shown in Fig.~\ref{fig:simelecdens}, the filaments develop within femtoseconds of the laser interacting with the target as can be seen in videos of the simulation included in the supplement. First the laser irradiates the spot, creating an ionization patch the shape of the laser projected onto the target. After approximately 30~fs, a well defined ionization front propagates at $0.77c$ radially away from the spot, where $c$ is the speed of light. There is a variation in the free electron density behind this front as a function of position in the thickness of the target, in which the back surface and front surface of the target are higher density ($\sim\invccmath{22}$) with the center of the target well below the critical density of the probe which is \invcc[6.9]{21}. Behind this front, another region of higher density expands outward radially at a slower speed of about $0.34c$. This high density region is characterized by the protons being ionized and the carbon and oxygen atoms ionized to the K-shell.  On the edge of this region, the filaments are found to form. The filaments are radially directed, in agreement with the experiment, with random pitch to the filaments through the thickness of the target, as shown in Fig.~\ref{fig:sim3d}~and~\ref{fig:simdownxlayer}. While an elementary calculation of the attenuation of the probe due to the initial ionization of the back surface is negligible, and thus might not be observable in shadowgraphy, the subsequent expansion of the high density region in the simulations is found to agree with the earliest observed speed of the ionization front in the experiment, which we measure in the vertical direction to be $0.4c$.

\subsection*{Ionization Variation in the Target}
In addition to the most stark feature of the filaments, there is a variation in the evolution of free electron density through the thickness of the target, which is shown in Fig.~\protect\ref{fig:moresim} and along the $x$-axis as shown in the inset of that figure. The variation can be classified by considering layers of the target through the thickness of the target along the $x$-axis, which are well resolved given the 10~nm cell sizes utilized in this simulation. The general trend is the front and back surfaces are more ionized compared to bulk thickness of the target, as shown in \protect\subref{fig:simdownxlayer}, with \protect\subref{fig:simbacklayer} showing the back is ionized to a higher degree. As previously discussed, outside of the initial irradiation of the spot, the density behind the fast ionization front is less than the critical density of the probe and thus difficult to observe experimentally.

There does exist some structure in the ionization as a function of radius and $x$ position beyond the general trend of ionization of the surfaces. One feature found from inspection of Fig.~\ref{fig:moresim} is on the front side shown in \protect\subref{fig:simfrontlayer} at a radius just beyond the filaments, the ionization is low but then increases at a larger radius. This is both shown in \protect\subref{fig:simfrontlayer} as the lower density shadow just on the right of the filaments and in the plot of the density through the $x$-axis \protect\subref{fig:simdownxlayer} by examining the volume between the dotted lines at $-0.23$~\um{} and $-0.1$~\um. Furthermore, the plot of the front layer \protect\subref{fig:simfrontlayer} shows that this feature has an ellipsoidal symmetry in the target plane. Upon further examination of Fig.~\ref{fig:simdownxlayer}, there is also a general higher density layer between $x=0.0$~\um{} and $x=0.1$~\um{} near the back of the target which only extends to a radius of 17~\um{}. When viewed as a layer in a manner similar to \protect\subref{fig:simbacklayer} and \protect\subref{fig:simfrontlayer} but not shown here, it appears as a disc. Finally, the back side of the target shown in Fig.~\ref{fig:simbacklayer} is almost completely ionized to a high degree, as discussed in the previous paragraph, with the highest ionization being the filaments and the shadow of the spot. Most of the ionization shown in the in \protect\subref{fig:simbacklayer} is due to the ionization in the last few cells of the target region being highlighly ionized.

\subsection*{Comparison of Simulations and the Experiment}
As noted previously the filamentation in the simulation occurs almost immediately while the laser is still reflecting off the target. However, in the experimental shadowgraphy, the filaments become apparent well after this, nearing a picosecond after the initial RLPI. One potential explanation is that the instability is seeded as the simulations suggest, during the laser interaction, however the very initial ionization is not directly observable experimentally. It is valuable to remind the reader that the simulations only capture the first 100~fs after the laser reaches the target which is comparable to the 80~fs FWHM of the probe illumination, making it difficult for the very initial ionization process to be captured by the probe. More evidence that the instability is seeded early and not in subsequent transport of the hot electrons is the appearance of filamentary structures in the polarization plane very early, which we attribute to pre-pulse. The simulations do not have a pre-pulse, so do not exhibit these features. Finally, synthetic "shadowgraphy" from the simulation was created to compare to the experimental videos, and taking into account the attenuation using a simple Lambert-Beer's law using the density of the plasma in the simulation as well as convolving with the finite duration of the probe pulse, it was found that up until the end of the simulation the filaments do not appear and only the shadow of the laser spot is visible. Nonetheless, it is assumed that these initial filaments grow in size as the ionization patch grows, after which they become visible experimentally.

\subsection*{Impact of This Observation} Finally, it is worth speculating about the implications of this observation. Instabilities have been found to be present in a host of experiments with high intensity laser experiments (including RLPI), and we can add this particular flavor of instability, filamentation of the surface plasma due to anisotropy in the laser mode to the battery of instabilities one would expect in laser plasma interactions. In addition to instabilities like Rayleigh-Taylor and others, it is conceivable that this instability would particularly affect reduced mass targets \cite{Neumayer2009,Tresca2011} that are used for either ion acceleration or as candidates for warm dense matter studies, and this instability could lead to inhomogeneities in the heating of the latter. With regards to ion acceleration, the traditional Weibel filamentation along the direction of laser propagation \cite{Metzkes2014} affect the mode of the accelerated ions. The radial instability observed in this work is potentially responsible for the radial striations found in lower energy proton spatial distributions of TNSA expansions from 1 micron thick solid targets\cite{Wagner2016}, not typically present in the central and higher energy protons. Following this train of thought, it was in fact observed in the simulations that just in and around the laser spot does not display radial filamentation, as the radial filamentation only occurs when hot electrons begin to leave the laser spot. Therefore, one potential mitigation of this instability in reduced mass target is to ensure the target is about the size of the laser spot in area by defocusing the beam.

\section*{Methods}\label{natcomm:methods}
\subsection*{Experimental Setup}
The experiment, displayed in Fig.~\ref{fig:expsetup}, was performed with a heavily modified dual multi-pass amplifier Ti:Sapphire KMLabs Red Dragon laser system which generates \SI{40}{\femto\second} (FWHM) pulses at \SI{780}{\nano\meter} wavelength with up to \SI{5}{\milli\joule} of short pulse energy reaching the target, focused by an F/1 protected gold coated off axis paraboloid, creating a \SI{1.8}{\micro\meter} diameter (FWHM) spot. Pulses were focused at a 45 degree angle of incidence on to the target which was a freely suspended, sub-micron ($\sim$0.5\um{}) thin sheet of ethylene glycol formed by the off-center collision of two liquid jets\cite{Morrison2018}. Despite the use of a liquid target flowing at \SI{24}{\meter\per\second}, the vacuum chamber pressure was maintained to less than \SI{50}{\micro\bar} during operation. For this system the non-linear effects on the mode propagating through focus are not oberved below \SI{55}{\milli\bar}. The laser-target interaction was illuminated by means of a short pulse probe beam created from amplifying the longer wavelength portion of different pulses of a common oscillator then is frequency doubled to produce \SI{420}{\nano\meter} probe allowing the \SI{390}{\nano\meter} second harmonic emission from the RLPI to be filtered out as described in the Feister et. al\cite{Feister2014}. The use of a common oscillator allows for optical synchronization of the shadowgraphy with the main pulse. Utilizing the repetition rate, we develop time resolved movies of the target evolution in the following way: for a given time delay, ten acquisitions of shadowgraphs are averaged. Then, out of the ten, the shadowgraph closest in total value to the average is chosen as the shadowgraph for this delay. This process is repeated across a number of delays to create the time resolved movie. To acquire a movie in a reasonable length of time, and roughly follows the natural time scales of the dynamics, the delays are chose such that the time between frames increases cubically starting a picosecond after the initial interaction.

\subsection*{Simulations}
Due to the geometry of the target and the the ionization region, a three-dimensional PIC simulation was performed. This simulation was expensive, requiring 5632 cores, pushing well into the tera-FLOP regime in terms of peak possible performance. The simulation had a high spatial resolution of 10~nm cells through the thickness of the target, 25~nm and 50~nm along the axis along the polarization axis and along the axis normal to the polarization plane respectively, which allowed detailed observation of the RLPI. The target had a thickness of 460~nm with a height of 40~\um{} and width of 40~\um{} at a density of \invcc[1.08]{22}, to closely follow the actual target. In this simulation the laser was a 42~fs \Wcm[5]{18} pulse with a 1.87~\um{} spatial Gaussian radius incident at a 45 degree incidence angle to the target with $p$-polarization. The focus of the pulse was placed at the target surface.

The number of atoms per cell is 8 per species, however, given the high spatial resolution of the simulation, this is not considered to be too low compared to other high fidelity three-dimensional simulations (for example, see Sentoku~\textit{et~al.} 2002~\cite{Sentoku2002}). The starting temperature was room temperature (0.025 eV) with atomic species; given the species were started neutral (without a plasma), the regions of the target would not be susceptible to grid heating~\cite{Langdon1970} before ionization. The simulation employs a field ionization model that follows the PPT/ADK rate~\cite{PPT1966,ADK1986} and allows for Coulomb collisions once regions of the target are ionized that follow the Spitzer rate~\cite{Spitzer1963}.

The absorbing boundary conditions used~\cite{Mur1981} are not ideal and we observed reflection of an amount of laser light three orders down in intensity off these boundaries which re-intersect the target. Light transmitted through the target and is re-incident makes a negligible contribution to the free electron density as the intensity is much lower than the incident light. The significant front surface reflection intersects with the target away from the initial laser focus after the relativistic electron expansion has passed. This second interaction occurs after the filaments have formed and was not observed to impact their evolution at late times, particularly for the target ionization on the opposite side of the laser direction (negative horizontal dimension in Fig.~\ref{fig:simelecdens} and negative $y$-axis in the inset of Fig.~\ref{fig:moresim}).

\subsection*{Data availability statement}
Additional experimental data generated in this study and not already provided in the supplemental data will be given upon a reasonable request. The LSP code is not freely available, and access to it should be discussed with Voss Scientific, Albuquerque, NM. However, if one has access to the code, the input files to perform the simulations utilized here can be given upon reasonable request of the corresponding author.

\section*{Author contributions statement}
 W.M.R., E.A.C., and C.O. provided leadership and oversight for the experimental and simulation effort. J.T.M. and K.M.G. conducted the experiments and performed analysis, supported by K.D.F.'s experimental setup, maintenance, and automation without which this work is not possible, and G.K.N. performed the simulations and comparisons to the experimental result. All authors reviewed, edited, and contributed to the manuscript.

\section*{Additional information}
This work was performed at AFRL at WPAFB supported by the Air Force Office of Scientific Research under LRIR Project 17RQCOR504 managed by Dr. Riq Parra. Moreover, this research was performed while the author G.K.N held an NRC Research Associateship award at the Air Force Research Laboratory, Dayton, OH.

 \textbf{Competing interests}~The authors declare no competing financial interests. 


\bibliography{thebibdot}

\end{document}